\documentclass[twocolumn,showpacs,pra,superscriptaddress,longbibliography]{revtex4-1}
\usepackage{times}
\usepackage{amsmath}
\usepackage{amssymb}
\usepackage{graphicx}
\usepackage[unicode]{hyperref}
\hypersetup{
   unicode=true,          
   plainpages=false,
   colorlinks=true,       
   citecolor=blue,        
}
\usepackage[caption=false,position=top,singlelinecheck=off,justification=raggedright]{subfig}
\usepackage{color}
\usepackage{pst-node}

\def \b1{{\bf 1}}

\newcommand{\bea}{\begin{eqnarray}}

\newcommand{\eea}{\end{eqnarray}}

\newcommand{\beq}{\begin{equation}}

\newcommand{\eeq}{\end{equation}}

\begin{document}

\title{Time crystals in a shaken atom-cavity system}

\author{Jayson G. Cosme}
\affiliation{Zentrum f\"ur Optische Quantentechnologien and Institut f\"ur Laser-Physik, 
Universit\"at Hamburg, 22761 Hamburg, Germany}
\affiliation{The Hamburg Center for Ultrafast Imaging, Luruper Chaussee 149, Hamburg 22761, Germany}

\author{Jim Skulte}
\affiliation{Zentrum f\"ur Optische Quantentechnologien and Institut f\"ur Laser-Physik, 
Universit\"at Hamburg, 22761 Hamburg, Germany}

\author{Ludwig Mathey}
\affiliation{Zentrum f\"ur Optische Quantentechnologien and Institut f\"ur Laser-Physik, 
Universit\"at Hamburg, 22761 Hamburg, Germany}
\affiliation{The Hamburg Center for Ultrafast Imaging, Luruper Chaussee 149, Hamburg 22761, Germany}


\date{\today}
\begin{abstract}
We demonstrate the emergence of a time crystal of atoms in a high-finesse optical cavity driven by a phase-modulated transverse pump field, resulting in a shaken lattice.  
This shaken system exhibits macroscopic oscillations in the number of cavity photons and order parameters at noninteger multiples of the driving period, which signals the appearance of an incommensurate time crystal. 
The subharmonic oscillatory motion corresponds to dynamical switching between symmetry-broken states, which are nonequilibrium bond ordered density wave states. Employing a semiclassical phase-space representation for the driven-dissipative quantum dynamics, we confirm the rigidity and persistence of the time crystalline phase. We identify experimentally relevant parameter regimes for which the time crystal phase is long lived, and map out the dynamical phase diagram. We compare and contrast the incommensurate time crystal with the commensurate Dicke time crystal in the amplitude-modulated case.
\end{abstract}
\maketitle

\section{Introduction}

Time crystals are nonequilibrium ordered states that spontaneously break time translation symmetry \cite{Wilczek2012,Shapere2012,Sacha2018,Else2019}. While ground-state realizations of this order have been pointed out to be infeasible \cite{Bruno2013,Noz2013,Watanabe2015}, a natural environment for time crystallization are out-of-equilibrium scenarios found in periodically driven systems \cite{Sacha2015,Else2016,Yao2017,Khemani2016,Choi2017,Zhang2017,Khemani2017,Russomanno2017,Ho2017,
Else2017,Yu2018,Rovny2018,Barfknecht2018,Giergiel2018,Mizuta2018,Huang2018,Smits2018,Gong2018,Gambetta2018,Sullivan2018,Zhu2019,Buca2019,Lazarides2019,Heugel2019,Flicker2018,Dumitrescu2018,Peng2018,Giergiel2019,Zhao2019,Matus2019,Pizzi2019,
Autti2018} or time-translation invariant systems with dissipation \cite{Iemini2018,Tucker2018,Lledo2019,Kessler2019}.  
Spontaneous breaking of time translation symmetry is displayed in an observable $\hat{O}$ evolving at a temporal symmetry breaking period $T_\mathrm{B}$, i.e., $\langle \hat{O}(t)\rangle = \langle \hat{O}(t+T_{\mathrm{B}})\rangle$. Additional defining features of a TC  are persistence of the time-translation symmetry breaking (TTSB) for long times and robustness against small perturbations. 

A specific class of TC phase called discrete time crystals (DTC) occurs in periodically driven systems where temporal symmetry is discrete. In this case, subharmonic response manifests itself in the breaking of discrete time translation symmetry by observables oscillating at an integer multiple of the driving period $T$, i.e., $\langle \hat{O}(t)\rangle = \langle \hat{O}(t+nT)\rangle$ with $n\in\{2,3,4,\dots\}$. 
Furthermore, the system spontaneously orders at one of the relative phases $2 \pi j/n +\phi_0$, where $j \in \{0,1,\dots,n-1\}$, and $\phi_0$ is a state dependent phase. Theoretical and experimental works on the existence and understanding of the DTC phase in isolated \cite{Sacha2015,Else2016,Yao2017,Khemani2016,Choi2017,Zhang2017,Khemani2017,Russomanno2017,Ho2017,
Else2017,Yu2018,Rovny2018,Barfknecht2018,Giergiel2018,Mizuta2018,Huang2018,
Smits2018} and dissipative \cite{Gong2018,Gambetta2018,Sullivan2018,Zhu2019,Buca2019,Lazarides2019,Heugel2019} systems have been reported.  

\begin{figure}[!htb]
\centering
\includegraphics[width=1.0\columnwidth]{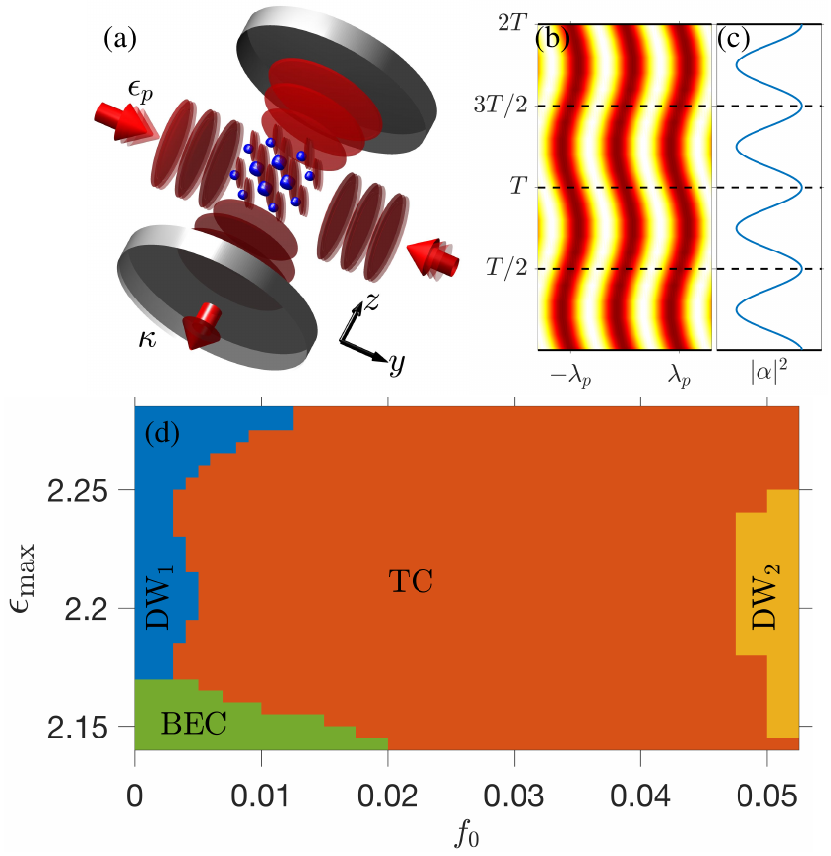}
\caption{(a) Schematic representation of the physical system. A gas of ultracold bosons in the density-wave phase is prepared inside a high-finesse cavity with a photon loss rate $\kappa$. The transverse pump beam with intensity $\epsilon_p$ and wavelength $\lambda_p$ is shaken at a period $T$. (b) Time evolution over two driving cycles of the standing-wave potential produced by the pump. (c) Steady-state response of the cavity mode occupation for nonresonant driving frequencies. The occupation follows the external driving with period $T/2$. (d) Dynamical phase diagram as a function of the maximum pump intensity $\epsilon_\mathrm{max}$ and unitless shaking amplitude $f_0$ for $\omega_d/2\pi=8.8~\mathrm{kHz}$. The TC phase is found to be stable for a wide region of the parameter space. }
\label{fig:schem} 
\end{figure}

The platform that we consider here are ultracold atoms in a high-finesse optical cavity  \cite{Ritsch2013}. Transverse pumping realizes the Dicke phase transition \cite{Dicke1954} between a homogeneous BEC phase and a self-organized density-wave (DW) order \cite{Domokos2002,Nagy2008,Baumann2010,Klinder2015}. 
In the following we propose to drive the system via lattice shaking; see Fig.~\ref{fig:schem}. This shaking method has proven to generate rich dynamical features in lattice systems with contact interactions \cite{Eckardt2017}. Here, we  broaden the scope of this method of dynamical control, by applying it to light-matter hybrid systems, in particular cavity-BEC systems, in which the photon field induces an infinite range interaction between the atoms. A shaken cavity-BEC system has also been studied in Ref.~\cite{Zhang2018}.

In this work, we propose to induce an \textit{incommensurate} time crystalline behavior in a dissipative system. To this end, we study the dynamics of atoms inside an optical resonator when the standing wave potential produced by a retroreflected pump beam in the transverse direction is periodically shaken.  Incommensurate time crystals  have dominant subharmonic response at a noninteger multiple of the drive, meaning $\langle \hat{O}(t)\rangle = \langle \hat{O}(t+mT)\rangle$ with a noninteger $m>1$ \cite{Flicker2018,Dumitrescu2018,Peng2018,Giergiel2019,Zhao2019,Matus2019,Pizzi2019,Autti2018}. 
In Fig.~\ref{fig:schem}(d) we summarize our results. We depict the dynamical regimes as a function of the pump intensity $\epsilon_p$ and the shaking amplitude $f_0$. For $f_0=0$, the system undergoes a phase transition at a critical pump intensity from a condensed state, indicated as BEC, to a density ordered state DW$_1$. For nonzero shaking amplitudes, the system forms a TC state. It emerges at the critical point, and covers a large region of the depicted parameter space. For large shaking amplitudes, this state competes with a striped density order, indicated as DW$_2$. In this paper, we describe the properties of the time crystalline state and the necessary shaking protocol to induce this state. This shaken system provides a natural platform for \textit{in situ} observation of a TC since the number of photons emitted from the cavity is monitored by a photodetector in experiments \cite{Baumann2010,Klinder2015,Klinder2016,Kessler2016}. The physical parameters used in this work are  motivated by the experimental setup in Refs.~\cite{Klinder2015,Klinder2015b,Klinder2016,Kessler2016,Georges2018} and thus, our predictions are experimentally relevant.  

This work is organized as follows. In Sec.~\ref{sec:system}, we describe the shaking protocol, the Hamiltonian, and the corresponding equations of motion for the atom-cavity system. 
The nonequilibrium states participating in the TTSB dynamics are presented in Sec.~\ref{sec:dbdw}. The dynamics of the order parameters describing the space-translation symmetry breaking is discussed in Sec.~\ref{sec:dyna}. Then in Sec.~\ref{sec:rob}, we further test the persistence and rigidity of the subharmonic response. We map out the dynamical phase diagram in Sec.~\ref{sec:dpd}. In Sec~\ref{sec:compare}, we compare the incommensurate TC in the shaken system with the commensurate or DTC in other driving schemes for atom-cavity systems. We demonstrate the robustness of the incommensurate TC in Sec.~\ref{sec:rob}.

 \section{System}\label{sec:system}

In the driven-dissipative system depicted in Fig.~\ref{fig:schem}(a), the pump beam is aligned along the $y$ direction and the cavity axis along the $z$ direction. 
The transverse pump mode is slowly and periodically shaken by introducing a time-dependent phase delay 
\begin{equation}
\varphi(t) = f_0 \mathrm{sin}(\omega_d t),
\end{equation}
which can be experimentally realized by using two optical modulators \cite{Zhang2018}. This produces a standing-wave potential $\mathrm{cos}(ky+\varphi(t))$, where $k=2\pi/\lambda_p$ and $\lambda_p$ is the pump wavelength, that is shaking at a  period of $T=2\pi/\omega_d$.
In the following discussion, we will work in a frame rotating at the frequency of the pump field $\omega_p$ \cite{Ritsch2013}. The Hamiltonian for the combined system consisting of the cavity, atomic, and the atom-field interaction reads
\begin{equation}\label{eq:2ham}
\hat{H} = \hat{H}_\mathrm{C} + \hat{H}_\mathrm{A} +\hat{H}_\mathrm{AC}.
\end{equation}
In Eq.~\eqref{eq:2ham}, the cavity dynamics with a single mode function $\mathrm{cos}(kz)$ is described by the Hamiltonian
\begin{equation}
 \hat{H}_\mathrm{C} = -\delta_{\mathrm{C}} \hat{\alpha}^{\dagger}\hat{\alpha},
\end{equation}
where $\delta_C < 0$ is the detuning between the pump and the cavity frequency and  $\hat{\alpha}$ ($\hat{\alpha}^{\dagger}$) is the cavity mode annihilation (creation) operator. We consider a two-dimensional system with atomic motion along the cavity axis and the transverse direction. The atomic part of the Hamiltonian is given by
\begin{equation}
\hat{H}_\mathrm{A} =\int dy dz \hat{\Psi}^{\dagger}(y,z)\left[-\frac{\hbar^2}{2m}\nabla^2 + V_{\mathrm{ext}}(y,z)\right]\hat{\Psi}(y,z),
\end{equation}
neglecting collisional atomic interaction.
The external potential is due to the standing wave created by retroreflecting the transverse pump beam, and is given by $V_{\mathrm{ext}}(y,z) = \omega_{\mathrm{rec}}\epsilon_p\mathrm{cos}^2(ky+\varphi(t))$. Finally, the atom-field interaction term is described by
\begin{align}
&\hat{H}_\mathrm{AC} =\int dy dz \hat{\Psi}^{\dagger}(y,z)\biggl[\Delta_0 \mathrm{cos}^2(kz)\hat{\alpha}^{\dagger}\hat{\alpha} \\ \nonumber
&+ \sqrt{\omega_{\mathrm{rec}}|\Delta_0|\epsilon_p }\mathrm{cos}(ky+\varphi(t))\mathrm{cos}(kz)(\hat{\alpha}^{\dagger}+\hat{\alpha}) \biggr]\hat{\Psi}(y,z).
\end{align}
Here, $\Delta_0<0$ is the light shift per intracavity photon and $\epsilon_p$ is the pump field intensity.

The spatial translation symmetry in the system motivates an expansion of the atomic field operator in the basis of plane waves 
\begin{equation}
\hat{\Psi}^{\dagger}(y,z) = \sum_{n,m}\hat{\phi}^{\dagger}_{n,m}\mathrm{e}^{i n k y}\mathrm{e}^{i m k z},
\end{equation}
where $\hat{\phi}_{n,m}$ ($\hat{\phi}^{\dagger}_{n,m}$) is the bosonic annihilation (creation) operator of single-particle modes corresponding to momentum $(n,m)\hbar k$.
This choice of basis set leads to the following Hamiltonian 
\begin{align}\label{eq:ham}
&\hat{H}(t)=-\delta_{\mathrm{C}}\hat{\alpha}^{\dagger}\hat{\alpha}+\frac{\Delta_0}{4}\hat{\alpha}^{\dagger}\hat{\alpha}\hat{Z} +\frac{\Delta_0}{2}\hat{\alpha}^{\dagger}\hat{\alpha}\hat{N} +\omega_{\mathrm{rec}}\hat{E}\\ \nonumber
& -\frac{\omega_{\mathrm{rec}}}{2}\epsilon_p\hat{N} - \frac{\omega_{\mathrm{rec}}}{4}\epsilon_p\hat{Y}(t)+\frac{\sqrt{\omega_{\mathrm{rec}}|\Delta_0|\epsilon_p}}{4}\hat{D}\hat{J}(t)
\end{align}
where $\hat{Z}=\sum \left( \hat{\phi}^{\dagger}_{n,m+2}\hat{\phi}_{n,m} + \mathrm{h.c.}\right)$, $\hat{N}=\sum \hat{\phi}^{\dagger}_{n,m}\hat{\phi}_{n,m}$, $\hat{E}=\sum (n^2+m^2)\hat{\phi}^{\dagger}_{n,m}\hat{\phi}_{n,m}$, and $\hat{D}=\hat{\alpha}^{\dagger}+\hat{\alpha}$. The time-dependent shaking is manifested in Eq.~\eqref{eq:ham} through $\hat{Y}(t)=\sum \left( \hat{\phi}^{\dagger}_{n+2,m}\hat{\phi}_{n,m}e^{i2\varphi(t)} + \mathrm{h.c.}\right)$ and  $\hat{J}(t)=\sum \left( \hat{\phi}^{\dagger}_{n,m}\left(\hat{\phi}_{n+1,m+1} + \hat{\phi}_{n+1,m-1} \right)e^{i\varphi(t)} + \mathrm{h.c.}\right)$. This Hamiltonian has a discrete time-translation symmetry $H(t) = H(t+T)$ due to the periodicity of the driving induced by the time-dependent phase $\varphi(t)$.

Ordered phases of the system can be identified via the relevant order parameters. The superradiant phase of the undriven system, in which the atoms self-organize at the antinodes of the emergent optical lattice, is described by the following order parameter \cite{Nagy2008,Baumann2010}
\begin{equation}\label{eq:opdw}
\Phi_{\mathrm{DW}_1} = \langle \mathrm{cos}(ky)\mathrm{cos}(kz) \rangle.
\end{equation}
When the strength of the pump field is modulated, other types of density wave ordering can be excited if the driving frequency matches the fundamental momentum excitations of the atomic ensemble \cite{Cosme2018}. For instance, an emergent dynamical striped phase may arise characterized by large amplitude oscillations of the order parameter, which is
\begin{equation}\label{eq:opdw4}
\Phi_{\mathrm{DW}_2} = \langle \mathrm{cos}(2ky) \rangle.
\end{equation} 
A new type of dynamical ordered state can emerge in which atoms self-organize at the nodes of the emergent optical lattice. Such phase is reminiscent of bond ordered states in solid and thus, we will refer to this phase as the bond density wave (BDW) state.
The order parameter for a BDW$_1$ state can be formulated by shifting the pump mode function by half of the pump wavelength $\lambda_p$:
\begin{equation}\label{eq:opbdw}
\Phi_{\mathrm{BDW}_1} = \langle \mathrm{sin}(ky)\mathrm{cos}(kz) \rangle.
\end{equation}
Similar to the DW$_1$ phase, the BDW$_1$ phase breaks the $\mathbb{Z}_2$ symmetry by self-organizing into one of the two possible checkerboard configurations. However, unlike the DW$_1$ phase, the atoms in the BDW$_1$ phase self-organize between the antinodes or ``bonds" of the emergent optical lattice. In the undriven system, the bonds are unstable positions for the atoms, and therefore this ordered state must be dynamical similar to other instabilities found in atom-cavity systems \cite{Keeling2010, Bhaseen2012, Piazza2015, Kessler2019, Chiacchio2019, Dogra2019}.

To capture how these bond orders couple to the driving, and the competition of orders, we determine the lowest-order terms of the free energy. Specifically, we assume a product of coherent states for the DW$_1$, DW$_2$, and BDW$_1$ operators, as well as for the photon mode. Taking the expectation value of the Hamiltonian of this state gives 
\begin{align}\label{eq:ham2}
&E(\Phi_{\alpha},\Phi_{\mathrm{DW}_1},\Phi_{\mathrm{BDW}_1},\Phi_{\mathrm{DW}_2}) \approx \\ \nonumber
& (s_1+s_2 +\nu_1|\Phi_{\alpha}|^2)\biggl(|\Phi_{\mathrm{DW}_1}|^2+|\Phi_{\mathrm{BDW}_1}|^2\biggr) \\ \nonumber
& +\frac{s_1}{2} \Phi_{\mathrm{DW}_2}\mathrm{cos}(2\varphi) + s_3|\Phi_{\alpha}|^2 \\ \nonumber
&+\nu_2\Phi_{\alpha,r}\biggl(\Phi_{\mathrm{DW}_1}\mathrm{cos}(\varphi) - \Phi_{\mathrm{BDW}_1}\mathrm{sin}(\varphi)\biggr)
\end{align}
to lowest order, where $s_1=-{\omega_{\mathrm{rec}}\epsilon_p}$, $s_2=4\omega_{\mathrm{rec}}$, $s_3=-\delta_{\mathrm{C}}$, $\nu_1=\Delta_0$, $\nu_2={\sqrt{\omega_{\mathrm{rec}}|\Delta_0|\epsilon_p}}$, $\Phi_{\alpha,r}=\mathfrak{R}\Phi_{\alpha}$, and $\varphi \equiv \varphi(t)$.
Given the small driving amplitudes $f_0 \ll 1$ that we consider here, we have $\mathrm{sin}(\varphi) \approx \varphi$, resulting in linear driving terms. Furthermore, the external driving couples parametrically to the product of the BDW$_1$ and the real part of the photon order parameter.
While DW$_1$ is the dominant equilibrium order for $\varphi(t)=0$ and for sufficiently large $\nu_2$, BDW$_1$ can only become the steady-state order in the presence of periodic driving. This requires a parametric resonance for BDW$_1$; see also Refs.~\cite{Chitra2015,Molignini2018}. At these resonances this nonequilibrium ordered state emerges, which breaks time-translation symmetry. Away from these resonances the spatially modulated density of the atoms oscillates around its equilibrium order with a small amplitude, as shown in Fig.~\ref{fig:schem}(c). 

We simulate the dynamics of the system in the semiclassical limit by solving the set of coupled equations of motion for the $c$ numbers of the momentum modes $\phi_{n,m}$ and the cavity mode $\alpha$ given by
\begin{align}\label{eq:eom}
i\frac{\partial {\phi}_{n,m}}{\partial t} &= \frac{\partial {H}(t)}{\partial {\phi}^{*}_{n,m}}\\ \nonumber
i\frac{\partial {\alpha}}{\partial t} &= \frac{\partial {H}(t)}{\partial {\alpha}^{*}}-i\kappa{\alpha} + i\xi.
\end{align}
Note that, in Eq.~\eqref{eq:eom}, we have included the decay term proportional to $\kappa$ in the cavity mode dynamics, which captures the rate at which photons are leaking out of the cavity, and the corresponding stochastic noise term $\xi(t)$ with $\langle \xi^*(t)\xi(t') \rangle = \kappa \delta(t-t')$ \cite{Ritsch2013}. 
Solving Eq.~\eqref{eq:eom} amounts to obtaining results in the thermodynamic limit as the mean-field approximation becomes exact when the number of atoms $N_a \to \infty$, while $N_a \Delta_0$ is kept constant \cite{Nagy2008,Gong2018}. This can be seen from how the equations of motion remain unaffected by rescaling $ {\alpha} \to  \sqrt{N_a}{\alpha}$ and $ {\phi_{n,m}} \to  \sqrt{N_a}{\phi_{n,m}}$  for $N_a \Delta_0=\mathrm{const}$.
Moreover, the equation of motion for the atoms shown in Eq.~\eqref{eq:eom} corresponds to a saddle point of the action after an expansion in powers of $1/N_a$, which further justifies the mean-field treatment as $N_a \to \infty$  \cite{Polkovnikov2010,Carusotto2013}.

\section{Time crystal Phase}\label{sec:TC}

We determine the dynamical response that is induced by the shaking process by solving Eq.~\eqref{eq:eom} numerically.
We use experimentally relevant parameters for $N_a = 60\times 10^3$, $\omega_{\mathrm{rec}}=2\pi \times 3.55~\mathrm{kHz}$, $\kappa=2\pi \times 4.50~\mathrm{kHz}$, and $\Delta_0= - 2\pi \times 0.36~\mathrm{Hz}$ based on the set up in Ref.~\cite{Klinder2015}. We choose a negative effective detuning $\delta_{\mathrm{eff}} \equiv \delta_C - (1/2)N_a\Delta_0=-2 \pi \times 22~\mathrm{kHz}$, where $N_a$ is the number of atoms. The system is operating in the recoil-resolved regime since $\kappa/4\omega_{\mathrm{rec}} < 1$, which means that a few-mode description is valid and the dynamics of the cavity and atomic modes happen on a comparable timescale \cite{Klinder2016}. This implies both equations of motion have to be solved on equal footing, and that the cavity mode in Eq.~\eqref{eq:eom} cannot be integrated out. When solving Eq.~\eqref{eq:eom}, we have used single-particle momentum modes spanning $\{n,m\}\in [-6,6]~\hbar k$ in order to guarantee the convergence of our results. Note that in the mean-field (MF) limit, we neglect the noise term in the cavity dynamics. However, this noise term must be included for calculations beyond the MF approximation in order to correctly treat the photon loss of the cavity, as will be done later in this work.

We initialize the system by ramping up the pump power linearly up to $\epsilon_{\mathrm{max}}$ starting from a BEC state. 
We then hold the pump power constant at $\epsilon_{\mathrm{max}}$ so that the system reaches a steady DW ordered state. Afterwards, the shaking protocol for the transverse standing wave is switched on. We display the time axis both in units of $T$ and in milliseconds (ms).

\subsection{Dynamical bond-density wave phase}\label{sec:dbdw}

\begin{widetext}

\begin{figure}[!htb]
\centering
\includegraphics[width=1.0\columnwidth]{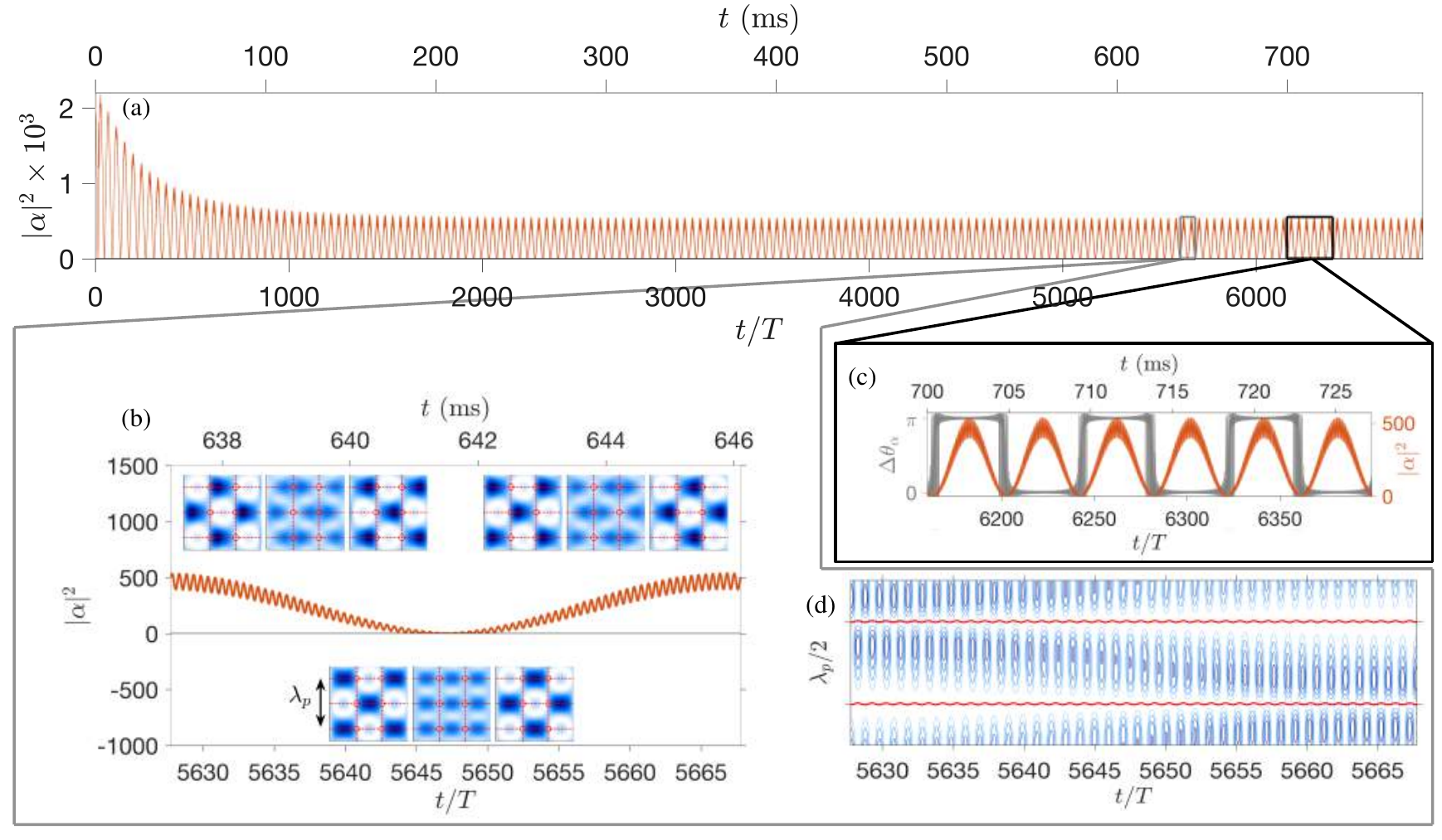}
\caption{(a) Cavity mode dynamics for $\omega_d/2\pi=8.8~\mathrm{kHz}$ with fixed $\epsilon_{\mathrm{max}}/E_{\mathrm{rec}}=2.24$ and $f_0=0.03$. (b) Dynamics of the BDW$_1$ phase. Insets: each set of inset plots depicts the evolution of the single-particle density over half a period of the driving cycle, i.e., $\rho(y,z,t)$ for $t\in\{t_1,t_1+T/4,t_1+T/2\}$. $\rho(y,z,t)$ around the time when $|\alpha|^2$ is at (top left) a maximum, (bottom) a minimum, and (top right) the next maximum. Circles mark the position of the antinodes of the combined mode functions $\mathrm{cos}(ky+\varphi(t))\mathrm{cos}(kz)$. (c) Dynamics for (left axis) the phase difference between the pump and cavity light fields $\Delta \theta_{\alpha}$ and (right axis) the cavity mode occupation. (d) Contour plot for the dynamics of the slice of the SPD at $\rho(y,z=0,t)$. The system oscillates between symmetry broken BDW$_1$ states at a fast time scale corresponding to the driving frequency. The SPD develops an asymmetry on a much longer time scale leading to the TTSB. Oscillating curves correspond to the position of the antinodes of the cavity-pump potential. }
\label{fig:big} 
\end{figure} 
\end{widetext}

We find that for frequencies larger than the frequency associated with the critical coupling strength $\epsilon_{\mathrm{cr}}$ of the superradiant phase transition, $\omega_{\mathrm{res}}=2\omega_{\mathrm{rec}}\sqrt{\epsilon_{\mathrm{cr}}} \approx 2\pi \times 8.7~\mathrm{kHz}$, a unique nonequilibrium phase exhibiting macroscopically slow oscillations of the cavity mode occupation emerges in the system as shown in Fig.~\ref{fig:big}. 
This subharmonic response of the cavity mode breaks the time-translation symmetry of the driven Hamiltonian, which is one of the important hallmarks of a TC. An example of a TC phase in the shaken system is presented in Fig.~\ref{fig:big} for $\omega_d/2\pi=8.8~\mathrm{kHz}$. The long-lived stability of the TTSB is depicted in Fig.~\ref{fig:big}(a) as the cavity mode occupation $|\alpha|^2$ periodically switches between zero and a non-zero number.

In Fig.~\ref{fig:big}(c), we show the dynamics of the cavity mode occupation and the relative phase difference between the pump and cavity light fields $\Delta \theta_{\alpha}= \mathrm{arg}(\alpha)$ for the TC phase. This information is accessible in experiments using a balanced heterodyne detection scheme for measuring the relative time-phase between the pump laser and the cavity field in real-time \cite{Baumann2011,Kollar2017}. As shown in Fig.~\ref{fig:big}(c), the relative phase difference switches between 
zero and $\pi$ at a symmetry breaking period $T_{\mathrm{B}}$, which is the characteristic frequency of the dynamical phase. This demonstrates the periodic switching of the system between states with broken discrete spatial-translation symmetry. A similar phenomenon of dynamical switching between symmetry broken states has been predicted in Refs.~\cite{Chitra2015,Molignini2018,Smits2018,Mizuta2018,Cosme2018,Zhu2019}.

Motivated by the dynamical switching inferred from the dynamics of the relative time phase, we now proceed to identify the type of symmetry broken phase associated with the TC order by calculating the single-particle density (SPD) profiles of the atomic ensemble in the MF limit:
\begin{align}
\rho(y,z) &\equiv \langle \Psi^{\dagger}(y,z)\Psi(y,z) \rangle \\ \nonumber
&= \sum_{n,m,n',m'} \phi^{\dagger}_{n,m}\phi_{n',m'}e^{i(n-n')ky}e^{i(m-m')kz}.
\end{align}

We obtain the SPD at various instants of time for the TC phase as depicted in the insets of Fig.~\ref{fig:big}(b). These SPD profiles imply that the ordered phase is a bond ordered state. Thus, we refer to this symmetry broken phase as a bond-density wave or BDW$_1$ phase. This can be seen from the inset plots in Fig.~\ref{fig:big}(b), where the maxima of the SPD $\rho(y,z)$ localizes at the nodes of the effective single-particle potential formed by the product of the pump and cavity mode functions $V_{\mathrm{PC}}(y,z)=\mathrm{cos}(ky+\varphi(t))\mathrm{cos}(kz)$.

For a spatially symmetric self-organized state at the nodes of $V_{\mathrm{PC}}(y,z)$, i.e., $\rho(y,z)=\rho(-y,z)$, the system has an empty cavity mode since 
\begin{equation}\label{eq:symm}
\int dy dz V_{\mathrm{PC}}(y,z) \rho(y,z) \approx 0.
\end{equation}
This leads to a vanishing atom-cavity coupling responsible for the two-photon process of exciting atoms into higher momentum modes. Thus, the cavity mode population vanishes. This is realized at periodic instants of time when the cavity mode occupation vanishes as seen in the bottom inset of Fig.~\ref{fig:big}(b). The point in time when the cavity mode occupation is minimal occurs when the atoms self-organize in a square lattice configuration with $\lambda_p/2$ spatial periodicity. For this spatial distribution, the Bragg condition necessary for scattering photons into the cavity is not satisfied.  
However, we find that the atomic ensemble self-organize asymmetrically in the direction of the shaking pump field at other instances of the driving cycle. In this case, the SPD is not symmetric around the position of the nodes of the combined pump-cavity field, i.e., $\rho(y,z)\neq\rho(-y,z)$, and this allows for the scattering of photons into the cavity because the atom-cavity coupling is now finite:
\begin{equation}\label{eq:asymm}
\int dy dz V_{\mathrm{PC}}(y,z) \rho(y,z) > 0.
\end{equation}
Typical density profiles of asymmetrically self-organized atoms are shown in the upper insets of Fig.~\ref{fig:big}(b). This asymmetry occurs because the time-dependent terms periodically break the translation symmetry along the pump direction of the Hamiltonian in Eq.~\eqref{eq:ham}.
 
The dynamics of the SPD profiles demonstrates the existence of a slow and fast oscillatory response of the driven system. 
The fast oscillatory response constitutes rapid switching between even and odd BDW$_1$ states as depicted in the top inset of Fig.~\ref{fig:big}(b), where the SPD $\rho(y,z,t)$ is presented at $t\in\{t_1,t_1+T/4,t_1+T/2\}$. Furthermore, in Fig.~\ref{fig:big}(d), the dynamics over one characteristic period of the cavity mode occupation is depicted for a slice of the SPD at $\rho(y,z=0,t)$. 
The resonant nature of the TC phase is apparent in the strong response of the system compared to the relatively small shaking amplitude depicted in Fig.~\ref{fig:big}(d) compared to the off-resonant response in Fig.~\ref{fig:schem}(c). 
Figure~\ref{fig:big}(d) shows the slow oscillation of the peaks of $\rho(y,z,t)$ between the nodes of $V_{\mathrm{PC}}(y,z)$, which occurs on a much slower time scale than the underlying fast oscillations. When the SPD becomes completely symmetric, see bottom inset of Fig.~\ref{fig:big}(b), the cavity mode becomes approximately empty consistent with Eq.~\eqref{eq:symm}. At the points in time when the SPD has an asymmetric distribution, photons scatter into the cavity, resulting in a nonvanishing occupation of the cavity mode. One period in $|\alpha|^2$ is completed when an odd checkerboard distribution transforms into the corresponding even checkerboard distribution or vice versa, see top inset of Fig.~\ref{fig:big}(b). The dynamics of the entire system has a temporal periodicity of $T_{\mathrm{B}}$: 
\begin{equation}
\rho(y,z,t_1+T_{\mathrm{B}}) = \rho(y,z,t_1).
\end{equation}
We point out that the degree of asymmetry of the SPD profile as the atoms self-organize on the bonds of the optical lattice is quantified by the BDW$_1$ order parameter in Eq.~\eqref{eq:opbdw}, which is the overlap between the SPD and the spontaneously formed optical lattice.

\subsection{Dynamics of the cavity and the order parameters}\label{sec:dyna}

\begin{widetext}

\begin{figure}[!htb]
\centering
\includegraphics[width=1.0\columnwidth]{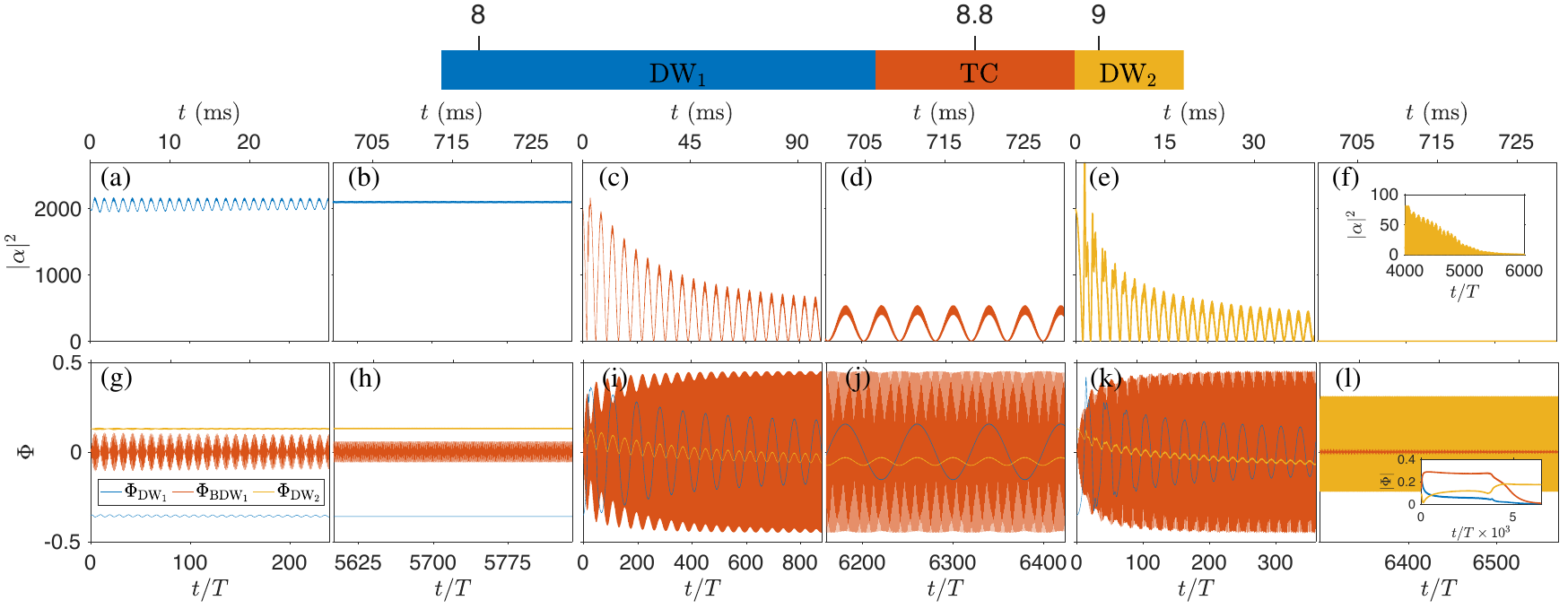}
\caption{Top: phase diagram as a function of $\omega_d/2\pi$ (in units of kHz) for fixed values of $\epsilon_{\mathrm{max}}/E_{\mathrm{rec}}=2.24$ and $f_0=0.03$. Bottom: exemplary dynamics of the (a)-(f) cavity mode $|\alpha|^2$ and (g)-(l) order parameters $\Phi_\mathrm{DW_1}$ (blue [dark gray]), $\Phi_\mathrm{BDW_1}$ (red [gray]), and $\Phi_\mathrm{DW_2}$  (yellow [light gray]) for the three dynamical phases: DW$_1$, TC, and DW$_2$.  (a),(b),(g),(h) $\omega_d/2\pi=8.0~\mathrm{kHz}$. (c),(d),(i),(j) $\omega_d/2\pi=8.8~\mathrm{kHz}$. (e),(f),(k),(l) $\omega_d/2\pi=9.0~\mathrm{kHz}$. Insets: (f) slow decay of the cavity mode occupation in the DW$_4$ phase; (l) slow decay of other order parameters $|\Phi|$ with Gaussian filtering except for $|\Phi_\mathrm{DW_4}|$ in the DW$_4$ phase. The Gaussian filter is chosen to remove the micromotion in $|\Phi|$.}
\label{fig:phases} 
\end{figure} 
\end{widetext}

We discuss the dynamics of the order parameters, in particular of $\Phi_{\mathrm{BDW}_1}$.
We map out the phase diagram for varying $\omega_d/2\pi \in [8.0,9.2]~\mathrm{kHz}$ and fixed pump power $\epsilon_{\mathrm{max}}/E_{\mathrm{rec}}=2.24$ and driving amplitude $f_0=0.03$ as shown in Fig.~\ref{fig:phases}. We also show exemplary dynamics for the cavity mode occupation and relevant order parameters in Fig.~\ref{fig:phases}. Three distinct phases are identified based on the leading order parameter and the long-time dynamics of the cavity mode. 

An example of the cavity mode dynamics for the DW$_1$ phase is shown in Figs.~\ref{fig:phases}(a) and \ref{fig:phases}(b). In the absence of shaking, the cavity mode has a constant, nonzero photon occupation. \cite{Domokos2002,Nagy2008,Baumann2010,Klinder2015}. This leads to a finite and dominant $\Phi_\mathrm{DW_1}$, as seen in Figs.~\ref{fig:phases}(g) and \ref{fig:phases}(h). This order parameter can be either negative or positive, which reflects the broken $\mathbb{Z}_2$ symmetry in the superradiant phase. For the shaken system considered here, the cavity mode occupation $|\alpha|^2$ oscillates at $2\omega_d$ for long timescales as seen from the inset of Fig.~\ref{fig:phases}(b). The frequency doubling in the dynamics of $|\alpha|^2$ is due to the orientation of the pump beam which is perpendicular to the cavity axis. As shown in Figs.~\ref{fig:schem}(b) and \ref{fig:schem}(c), $|\alpha|^2$ oscillates with $2\omega_d$.

For driving frequencies larger than the parametric resonant frequency, $\omega_{\mathrm{res}}/2\pi \approx 8.7 ~\mathrm{kHz}$, the dynamical BDW$_1$ phase emerges. This phase is exemplified by $\omega_d/2\pi=8.8~\mathrm{kHz}$ in Figs.~\ref{fig:phases}(c) and \ref{fig:phases}(d). 
The BDW$_1$ order parameter $\Phi_{\mathrm{BDW}_1}$ exhibits emergent oscillations $\omega_{\mathrm{BDW}_1}$ between positive and negative values close to the parametric resonant frequency $\omega_{\mathrm{BDW}_1} \approx \omega_{\mathrm{res}}$, i.e., $\Phi_{\mathrm{BDW}_1}(t)=\Phi_{\mathrm{BDW}_1}(t+{2\pi}/{\omega_{\mathrm{res}}})$. Moreover, $\Phi_{\mathrm{BDW}_1}$ displays the largest oscillation amplitude amongst the relevant order parameters as depicted in Figs.~\ref{fig:phases}(i) and \ref{fig:phases}(j). This observation is consistent with the dynamics of the SPD seen in Fig.~\ref{fig:big}(b). 
The DW$_1$ order parameter $\Phi_{\mathrm{DW}_1}$ oscillates at $\omega_{\mathrm{DW}_1}=\omega_d - \omega_{\mathrm{BDW}_1}$. This leads to extremely slow oscillations since $\omega_{\mathrm{DW}_1} \ll \omega_d $ for $\omega_d$ close to the resonant frequency. 
More importantly, this highlights the incommensurate nature of the subharmonic response
\begin{equation}\label{eq:tb}
T_{\mathrm{B}} = \frac{2\pi}{\omega_{\mathrm{DW}_1}} =  \frac{2\pi}{\omega_d}\biggl(1 - \frac{\omega_{\mathrm{res}}}{\omega_d}\biggr)^{-1} = \frac{2\pi}{\omega_d(1-m)}.
\end{equation}
Generally, the ratio between the resonant and driving frequencies $\omega_{\mathrm{res}}/\omega_d $ can be any real number $0<m<1$  since $\omega_{\mathrm{res}}$ is independent of the driving frequency and only depends on the microscropic parameters of the atom-cavity system. This implies that the TC can be fine-tuned to a commensurate TC as demonstrated in Fig.~\ref{fig:big}, where $T_{\mathrm{B}} = 80 T$.
Note that the characteristic frequency of $|\alpha|^2$ in the incommensurate TC is $\omega_{|\alpha|^2} = 2\omega_{\mathrm{DW}_1}$.

If we increase the driving frequency further away from $\omega_{\mathrm{res}}$, the TC phase becomes unstable as it \textit{melts} into a new nonequilibrium ordered phase. The cavity mode dynamics for this phase is shown in Figs.~\ref{fig:phases}(e) and \ref{fig:phases}(f). The cavity becomes empty for long times as seen in the inset of Fig.~\ref{fig:phases}(f) because of the selective excitation of momentum modes corresponding to momentum of $\pm 2~\hbar k$ along the direction of the transverse pump beam. 
The process of populating the $\{k_y,k_z\}=\{\pm 2,0\}~\hbar k$ momentum modes is evidenced by the oscillatory behavior of the $\Phi_{\mathrm{DW}_2}$ shown in Fig.~\ref{fig:phases}(l).
The absence of the two-photon scattering process of atoms into the $\pm 1~\hbar k$ momentum modes forbids the runaway process required for the self-organization of atoms into a density-wave phase. This dynamical phase is also characterized by the periodic switching of atoms between symmetry broken striped phases as the maxima of the density distribution of atoms are shifted by $\lambda_p/4$ periodically in time. 
This steady state of the DW$_2$ order is preceded by transient TC order; see Fig.~\ref{fig:phases}(e). 
In fact, as shown in the inset of Fig.~\ref{fig:phases}(l), the DW$_2$ phase appears to exhibit a metastable TTSB, which lasts for relatively long timescales up to 4000 driving cycles or 400 ms of experimental time. 
This timescale is already close to typical experimental times in atom-cavity experiments, after which unwanted atom losses and decoherence effects from the environment become more dominant. Note that the subharmonic frequency of the metastable TC phase seen in Figs.~\ref{fig:phases}(e) and \ref{fig:phases}(k) still follows $\omega_{\mathrm{B}}=(\omega_d - \omega_{\mathrm{BDW}_1})$. This suggests that the resonant excitation of the dynamical BDW$_1$ phase is responsible for the TTSB in the shaken system.

\subsection{Dynamical phase diagram}\label{sec:dpd}

We map out the dynamical phase diagram in Fig.~\ref{fig:schem}(d) as a function of the pump intensity and the driving amplitude for fixed driving frequency $\omega_d/2\pi=8.8~\mathrm{kHz}$. For weak driving amplitudes $f_0$, we recover the phase transition from a homogeneous BEC phase to a self-organized superradiant DW$_1$ phase. As we increase the driving amplitude, the DW$_1$ phase becomes dynamically unstable and the TC phase starts to emerge. The TC phase exists in a large region of parameter space depicted in Fig.~\ref{fig:schem}(d). For large $f_0$, the TC phase becomes unstable to DW$_2$ order as demonstrated in Figs.~\ref{fig:phases}(k) and  \ref{fig:phases}(l).

\begin{figure}[!htb]
\centering
\includegraphics[width=1.0\columnwidth]{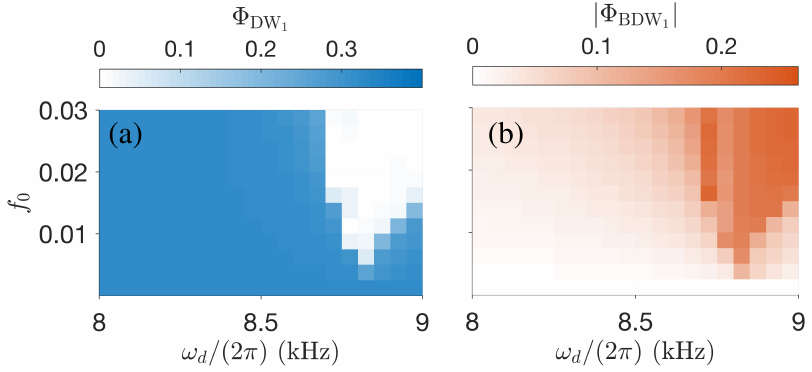}
\caption{Long-time averaged (a) $\Phi_{\mathrm{DW}_1}$ and (b) $|\Phi_{\mathrm{BDW}_1}|$ as a function of $\omega_d$ and $f_0$ for fixed $\epsilon_{\mathrm{max}}/E_{\mathrm{rec}}=2.24$ .}
\label{fig:stab} 
\end{figure}

In Fig.~\ref{fig:stab}, we present the long-time averaged values of the order parameters $\Phi_{\mathrm{DW}_1}$ and $|\Phi_{\mathrm{BDW}_1}|$ for fixed pump intensity and varying shaking amplitude $f_0$ and frequency $\omega_d$. The fluctuations in the initial momentum occupations are chosen to yield a positive order parameter in the equilibrium DW$_1$ phase. The emergence of oscillating order parameters leads to a vanishing long-time averaged $\Phi_{\mathrm{DW}_1}$ and a nonzero $|\Phi_{\mathrm{BDW}_1}|$, as indicated by the white region in Fig.~\ref{fig:stab}(a) and the dark region in Fig.~\ref{fig:stab}(b).

\begin{figure}[!htb]
\centering
\includegraphics[width=1.0\columnwidth]{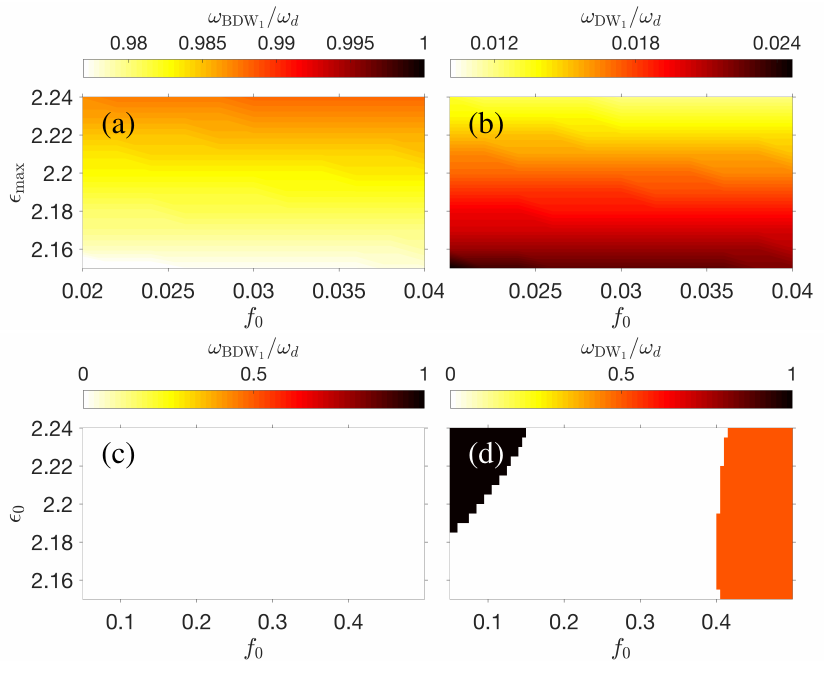}
\caption{Comparison between (a),(b) the dynamical bond-density wave phase and (c),(d) the dynamical normal phase. Dependence of the ratio (left) $\omega_{\mathrm{BDW}_1}/\omega_d$ and (right) $\omega_{\mathrm{DW}_1}/\omega_d$ for varying pump intensity and driving strength. For the dynamical BDW$_1$ phase, we use $\omega_d/2\pi = 8.8~\mathrm{kHz}$. For the dynamical normal phase, the amplitude of the pump field is modulated by $\omega_d/2\pi = 8.0~\mathrm{kHz}$.}
\label{fig:TCphased} 
\end{figure}

In Fig.~\ref{fig:TCphased}, we show the frequency response of the system as it changes with the pump intensity and the driving amplitude. We obtain the ratio between the emergent oscillation frequency of $\Phi_{\mathrm{BDW}_1}$ and the driving frequency, $\omega_{\mathrm{BDW}_1}/\omega_d$, over a time window $t\in [100, 200]~\mathrm{ms}$ and present them in Fig.~\ref{fig:TCphased}(a) for $\omega_d/2\pi=8.8~\mathrm{kHz}$. This frequency response $\omega_{\mathrm{BDW}_1}$ has a weak dependence on $f_0$ in particular for pump intensities close to the critical value for the superradiant phase transition. We find that the dependence of the emergent frequency $\omega_{\mathrm{BDW}_1}$ on the pump intensity is consistent with $\omega_{\mathrm{BDW}_1} \approx \omega_{\mathrm{res}} = 2\omega_{\mathrm{rec}}\sqrt{\epsilon_p}$; see Sec.~\ref{sec:dbdw}. The incommensurability of the subharmonic frequency is observed in Fig.~\ref{fig:TCphased}(b). 

\subsection{Comparison with Dicke time crystals}\label{sec:compare}

We compare the dynamical bond-density wave phase with the dynamical normal phase (DNP) first proposed in Ref.~\cite{Chitra2015} induced via amplitude modulation. The DNP is also considered as a Dicke time crystal in Refs.~\cite{Gong2018,Zhu2019,Else2019}. We will discuss the similarities and differences between the two nonequilibrium phases in the context of TTSB and their properties as TC phases in driven atom-cavity systems.

\begin{figure}[!htb]
\centering
\includegraphics[width=1.0\columnwidth]{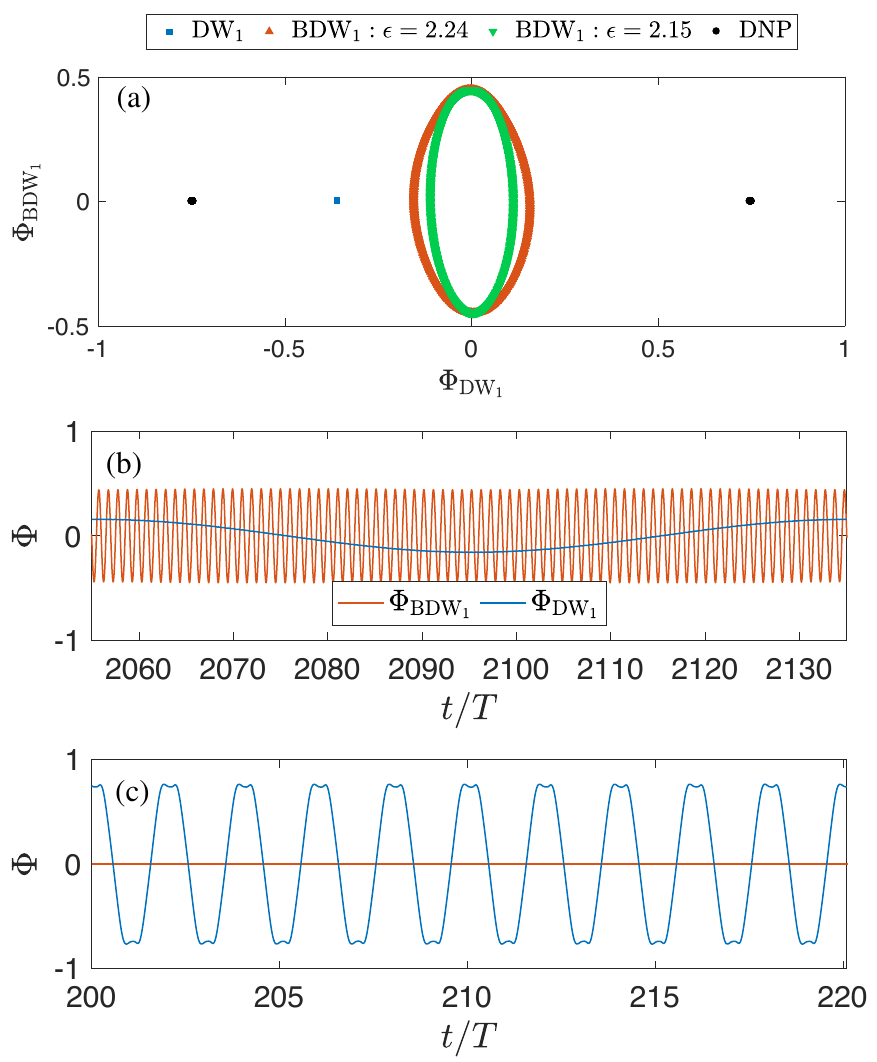}
\caption{(a) Stroboscopic order parameter portraits of $\Phi_{\mathrm{DW}_1}$ and $\Phi_{\mathrm{BDW}_1}$ for three phases: DW$_1$, BDW$_1$, and DNP. For the portraits, the relevant order parameters are recorded every $mT$ for integer $m$ in the long-time limit. Representative long-time dynamics of $\Phi_{\mathrm{DW}_1}$ (blue [dark gray]) and $\Phi_{\mathrm{BDW}_1}$ (red [gray]) for (b) the dynamical BDW$_1$ phase and (c) the DNP. }
\label{fig:TCphasesketch} 
\end{figure}

We obtain the values of the order parameters $\Phi_{\mathrm{DW}_1}$ and $\Phi_{\mathrm{BDW}_1}$ at integer multiples of the driving period $T$ to construct the stroboscopic order parameter portraits presented in Fig.~\ref{fig:TCphasesketch}(a). In addition to the DNP and dynamical BDW$_1$ phase, we also show an example for the standard DW$_1$ phase to emphasize the key differences between the three phases. For the DW$_1$ and BDW$_1$ phase, we use the same parameters as in Fig.~\ref{fig:phases}. For the DNP phase, we simulate the driven atom-cavity system with the same set of parameters but with amplitude modulation $\sqrt{\epsilon_p} = \sqrt{\epsilon_{0}}(1+f_0\mathrm{cos}(\omega_d t))$ as done in Ref.~\cite{Cosme2018}. For the amplitude-modulated case, we use a relatively strong driving strength $f_0=0.40$ in order to push the system beyond the regime of light-induced renormalization of the phase boundary \cite{Cosme2018,Georges2018} depicted as the dark region in Fig.~\ref{fig:TCphased}(d). 

The DNP is characterized by periodic switching of the DW$_1$ order parameter as the system switches from one of the two possible self-organized checkerboard patterns to the other \cite{Chitra2015,Molignini2018}. Such behavior is observed in the period-doubling dynamics of $\Phi_{\mathrm{DW}_1}$ in Figs.~\ref{fig:TCphased}(d) and \ref{fig:TCphasesketch}(c). 
The dynamical behavior of the DNP is similar to recently proposed TC phases which rely on coherent switching between symmetry broken states in space \cite{Smits2018,Mizuta2018,Lledo2019,Zhu2019}. For this type of TC phase, the subharmonic frequency is tied to the number of symmetry broken states participating in the dynamics of the system \cite{Mizuta2018}. In the case of the DNP, there are two kinds of possible $\textbf{Z}_2$-symmetry broken states leading to period doubling. Therefore, the DNP is a commensurate TC. This is demonstrated in Fig.~\ref{fig:TCphased}(d), where the DNP with $T_{\mathrm{DW}_1} = 2T$ arises for strong driving amplitudes. 

The dynamical BDW$_1$ phase exhibits time-dependent order parameters that break the time-translation symmetry as depicted in Fig.~\ref{fig:TCphasesketch}(b).
However, the dynamical BDW$_1$ phase has a tunable subharmonic frequency, in contrast to the period doubling dynamics of the DNP. 
For the DNP, the maxima of the atomic distribution are always at the antinodes of the light field leading to a fixed value of $\Phi_{\mathrm{BDW}_1}=0$ as seen in Figs.~\ref{fig:TCphased}(c) and \ref{fig:TCphasesketch}(a). 
For the BDW$_1$ phase, the density maxima oscillate around the antinodes. This breaks the parity symmetry along the pump direction. 
In Fig.~\ref{fig:TCphasesketch}, the atomic motion is represented in the plane spanned by the DW$_1$ and the BDW$_1$ order parameter. The lattice shaking protocol that we propose here expands the motion out of the on-lattice states. 

\section{Persistence and rigidity of time crystallinity}\label{sec:rob}

We now test the robustness of the TC phase against temporal perturbations and many-body correlations. 
To this end, we employ a semiclassical treatment of the dynamics based on the truncated Wigner approximation (TWA) \cite{Polkovnikov2010,Blakie2008}. The TWA is a semiclassical phase-space method that approximates the quantum dynamics by solving the equations of motion in Eq.~\eqref{eq:eom} for an ensemble of initial states which samples the initial Wigner distribution \cite{Polkovnikov2010,Blakie2008,Carusotto2013}. 
This method captures quantum aspects of the dynamics in the semiclassical limit by accounting for the leading-order quantum corrections to the MF solutions; for example, see \cite{Mathey2014,Cosme2014,Acevedo2017,Cosme2018a,Nagao2018,Kordas2013}.
Even in the absence of imperfections in the driving protocol, inherent perturbations that can destabilize or melt the time crystal are present in the system: (i) quantum fluctuations in the initial state and (ii) nonunitary imperfection in the time evolution captured by the stochastic noise $\xi$ associated to the dissipation of photons out of the cavity. 
The rigidity of TTSB dynamics to stochastic noise associated with the dissipation of photons suggests a nontrivial stabilization of a TC in dissipative systems \cite{Yao2017,Gong2018}.
We initialize the TWA simulations starting from a pure condensate state occupying the lowest momentum mode while other available but initially unoccupied cavity and atomic modes are populated with vacuum fluctuations. Then, we use a similar protocol for ramping into the DW state followed by the time-dependent shaking described in Sec.~\ref{sec:dyna}. We use $10^3$ trajectories to sample the initial quantum noise in the system.
\begin{figure}[!htb]
\centering
\includegraphics[width=1.0\columnwidth]{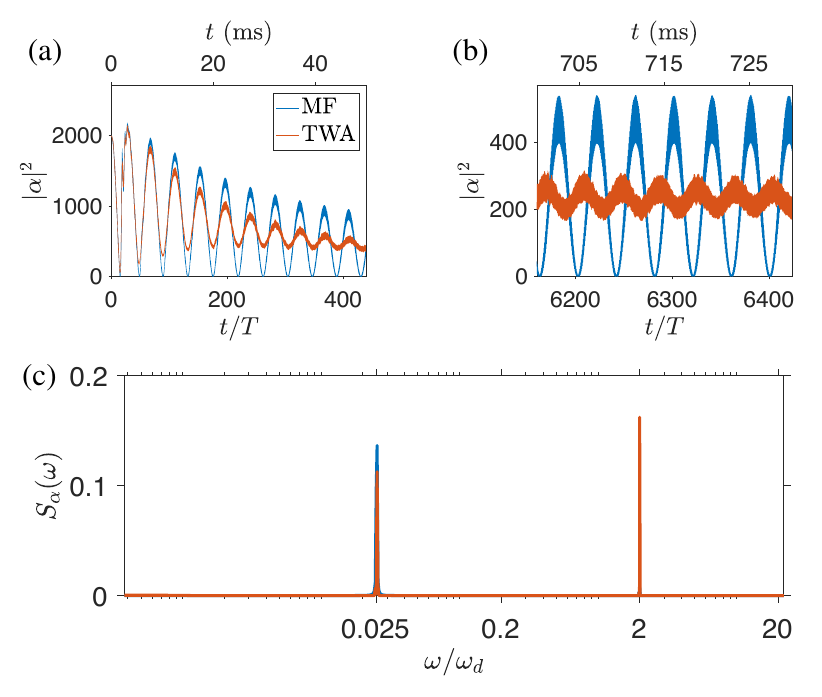}
\caption{Rigidity of TTSB to quantum fluctuations. MF (blue [dark gray]) and TWA (red [gray]) results for the cavity dynamics at (a) short and (b) long times. (c) Frequency spectrum for the cavity dynamics. The parameters are $\epsilon_{\mathrm{max}}/E_{\mathrm{rec}}=2.24$, $f_0=0.03$, and $\omega_d/2\pi=8.8~\mathrm{kHz}$.}
\label{fig:power} 
\end{figure} 
\begin{figure}[!htb]
\centering
\includegraphics[width=1.0\columnwidth]{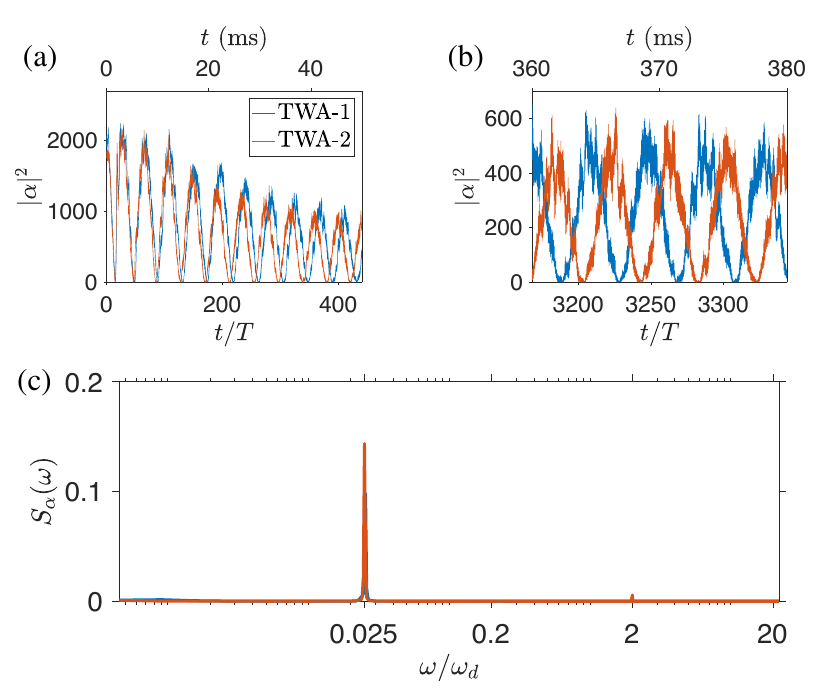}
\caption{Persistence of TTSB for two sample trajectories in TWA. Cavity dynamics for (a) short and (b) long timescales. (c) Fourier spectrum as in Fig.~\ref{fig:power}. The parameters are the same as Fig.~\ref{fig:power}.}
\label{fig:twpower} 
\end{figure} 
We calculate the frequency spectrum of the cavity dynamics $S_\alpha(\omega)$ by taking the Fourier transform of the cavity dynamics $\mathcal{F}\{|\alpha(t)|^2\}$ within the time window $t\in [400, 800]~\mathrm{ms}$. In Fig.~\ref{fig:power}(c), we show the results for $\omega_d/2\pi=8.8~\mathrm{kHz}$. The TWA simulation gives a similar peak as the MF simulation, which suggests that the dynamical state is robust against quantum fluctuations. We note that the TWA simulation displays an additional peak near the normal frequency response $\omega/\omega_d = 2$. Nevertheless, the relative photon number fluctuation predicted in Fig.~\ref{fig:power}(b) is still within the sensitivity of photodetectors in atom-cavity experiments. 

In Fig.~\ref{fig:twpower}, we show two TWA trajectories. We interpret these as single realizations of the atom cavity dynamics for highly occupied modes. This suggests that TTSB can be observed in a single experimental realization. The main peak observed in Fig.~\ref{fig:twpower}(c) for each trajectory corroborates the rigidity of the TC against temporal perturbations from the stochastic noise associated with the dissipation.

\begin{figure}[!htb]
\centering
\includegraphics[width=1.0\columnwidth]{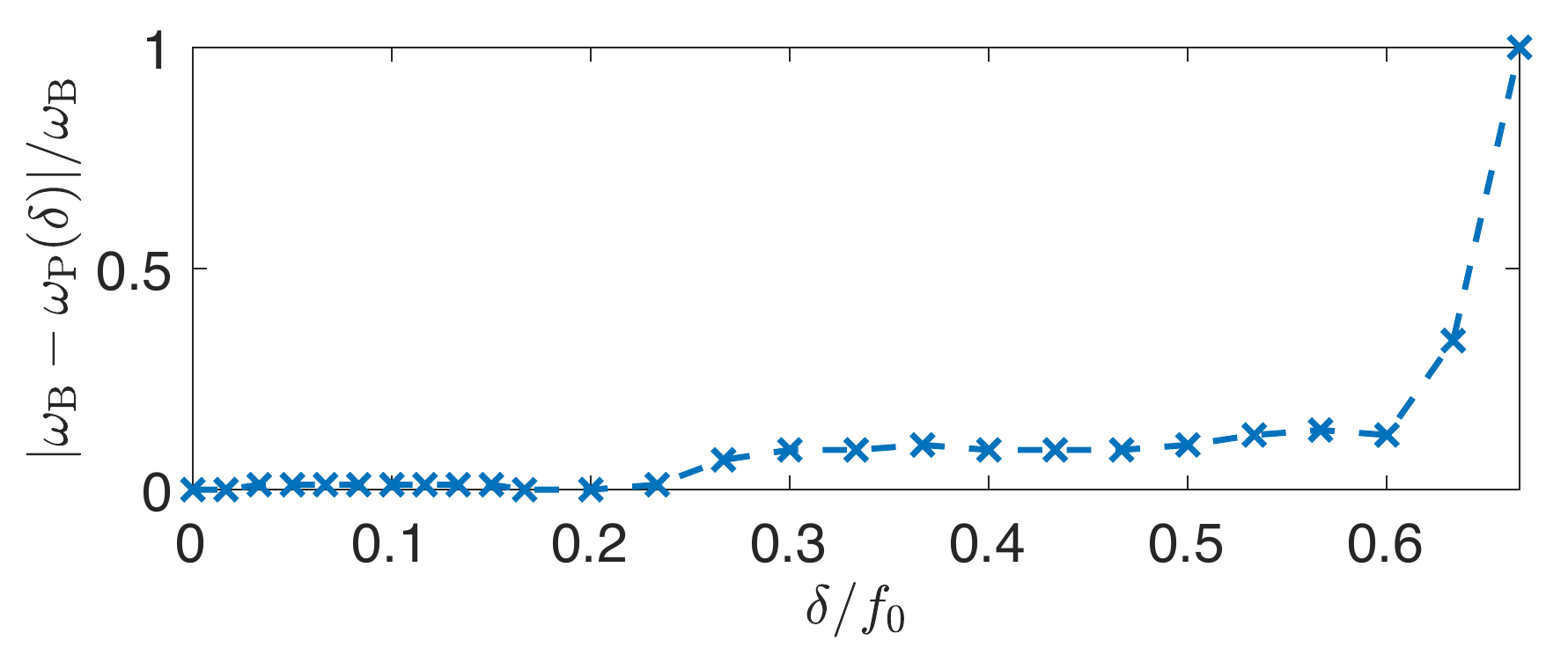}
\caption{Persistence of TTSB quantified by the relative shift in the frequency response in the presence of perturbation $\delta/f_0$ in the second-half of the driving. The parameters are the same as Fig.~\ref{fig:power}.}
\label{fig:pert} 
\end{figure} 
Next, we show the rigidity of $\omega_{\mathrm{B}}$ against imperfections of the driving protocol. We include temporal imperfections in the time-dependent Hamiltonian by introducing small perturbations in the parameters of the driving scheme. Here, we increase the amplitude of the shaking by an amount $\delta$ during the second half of a driving
\begin{equation}
\phi(t+T)=\phi(t) = \left\{
        \begin{array}{ll}
            f_0\mathrm{sin}(\omega_d t) & \quad 0\leq t < \frac{T}{2} \\
            (f_0+\delta)\mathrm{sin}(\omega_d t) & \quad \frac{T}{2} \leq t < T.
        \end{array}
    \right.
\end{equation}
We simulate the MF dynamics and obtain the corresponding subharmonic frequency response of the cavity mode $\omega_{\mathrm{P}}(\delta)$ of the perturbed system for varying $\delta$. The driven system is simulated until 1.6 s or roughly $14\times 10^3$ driving periods. The frequency peak is obtained from $t\in[1.2,1.6]$ s. The shift in the frequency response $|\omega_{\mathrm{B}}-\omega_{\mathrm{P}}(\delta)|/\omega_{\mathrm{B}}$ as a function of the relative strength of the perturbation $\delta/f_0$ is depicted in Fig.~\ref{fig:pert}. For weak imperfections in the driving $\delta/f_0 \lessapprox 20\%$, the time crystal displays the same periodicity.
This rigidity can be attributed to the long-range spatial ordering and the cavity-mediated all-to-all coupling of the atoms in the BDW phase, which protects the system from temporal perturbations.
For stronger perturbations, the time crystal becomes unstable as its periodicity shifts. The TC eventually melts into a typical DW phase for strong enough imperfections. 

\begin{figure}[!htb]
\centering
\includegraphics[width=1.0\columnwidth]{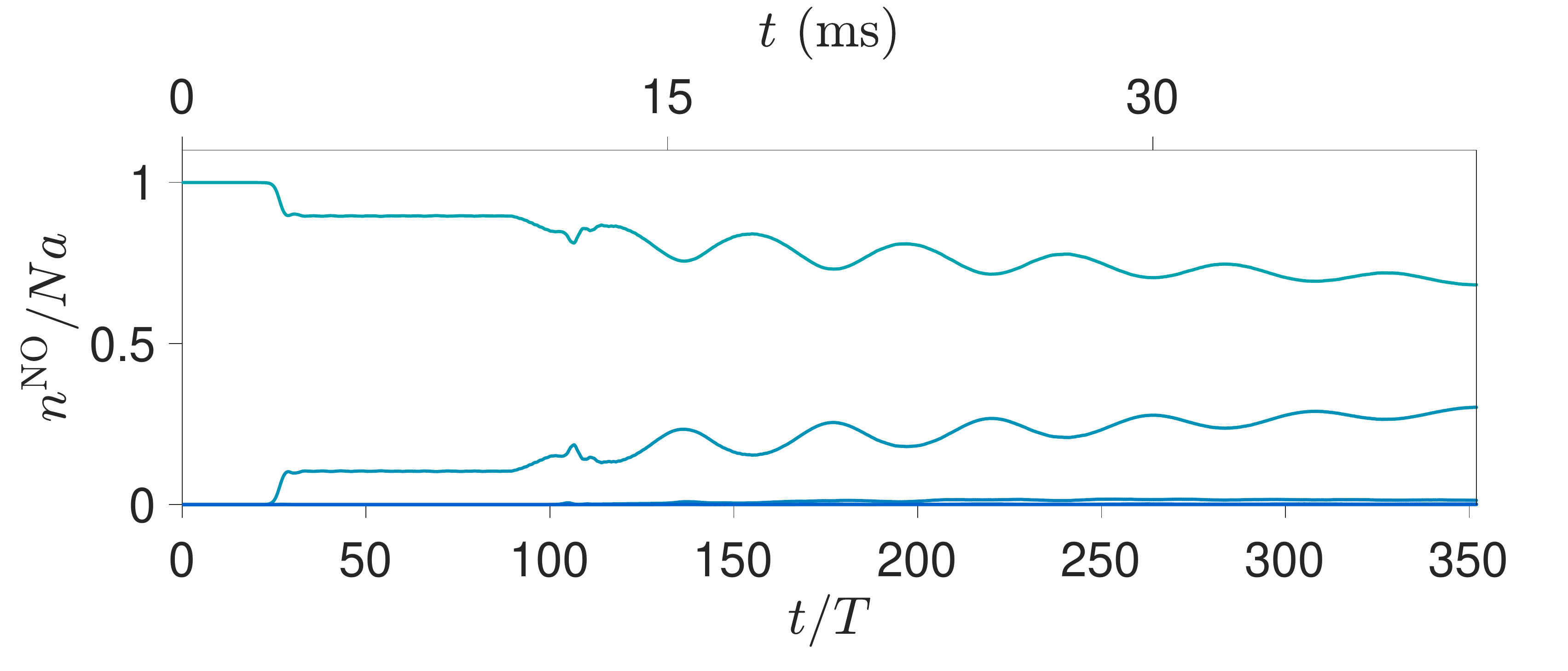}
\caption{Dynamics of the five largest eigenvalues (from light to dark gray) of the SPDM for the same parameters as in Fig.~\ref{fig:power}. Only two eigenmodes of the SPDM are highly occupied in the TC phase.}
\label{fig:mb} 
\end{figure} 
It is also important to demonstrate the robustness of the TC against many-body correlations especially for systems with all-to-all coupling \cite{Tucker2018,Zhu2019,Else2019}.  The leading-order quantum effects due to many-body correlation are captured within TWA. This can be quantified by the eigenvalues of the single-particle density matrix (SPDM), $\langle \Psi^{\dagger}(y',z') \Psi^{\dagger}(y,z) \rangle$. The SPDM is a quantity related to the Penrose-Onsager criterion for condensates in interacting systems and its largest eigenvalue corresponds to the condensate fraction \cite{Penrose1956}. Typical dynamics of the five lowest eigenvalues of the SPDM are shown in Fig.~\ref{fig:mb}. The fragmentation of the BEC is signaled by a decrease in the condensate fraction together with an increase in the remaining eigenvalues. Nevertheless, the corresponding cavity mode dynamics still exhibit TTSB as depicted in Fig.~\ref{fig:power}. This suggests that the dynamical BDW phase is stable beyond the mean-field regime similar to Dicke TC \cite{Zhu2019}. 

\begin{figure}[!htb]
\centering
\includegraphics[width=1.0\columnwidth]{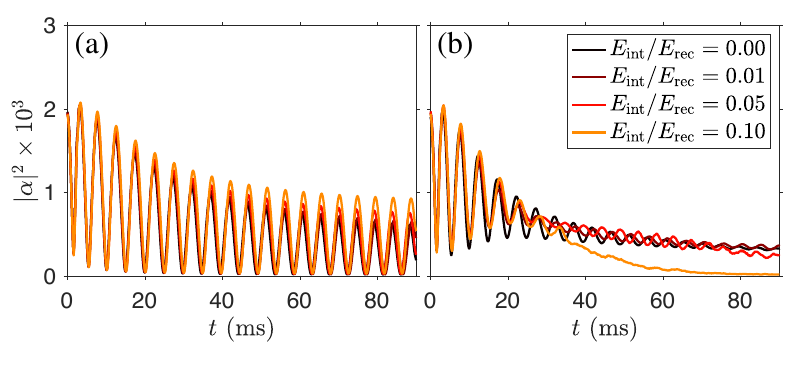}
\caption{(a) MF and (b) TWA results for dynamics of the cavity mode in the presence of collisional atom-atom interaction quantified by the interaction energy $E_\mathrm{int}$ (from black to yellow [light gray] for increasing $E_\mathrm{int}$). We filter out frequency faster $\pi/T_{\mathrm{B}}$ using a Gaussian filter. The shaking amplitude is fixed to $f_0=0.03$, while the pump intensity and the driving frequency are adjusted to match the initial $|\alpha|^2$ and $T_B$ for different $E_\mathrm{int}$.}
\label{fig:mbu} 
\end{figure} 
The effects of repulsive contact interaction between the atoms are taken into account by adding the term $\hat{H}_\mathrm{AA} =g_a \int dy dz \hat{\Psi}^{\dagger}(y,z)\hat{\Psi}^{\dagger}(y,z)\hat{\Psi}(y,z)\hat{\Psi}(y,z)$ to Eq.~\eqref{eq:2ham}.
The collisional interaction strength $g_a$ is measured in terms of the interaction energy $E_\mathrm{int}=N_a g_a/\lambda^2_p $ and its presence pushes the critical pump strength for the DW$_1$ transition to higher values \cite{Nagy2008}. Thus, in order to compare different $E_{\mathrm{int}}$, we tune $\epsilon_\mathrm{max}$ to match $|\alpha|^2$ in the DW$_1$ phase before shaking the pump. Morever, we adjust $\omega_d$ such that different $E_{\mathrm{int}}$ will yield the same $T_\mathrm{B}$ in the MF level as depicted in Fig.~\ref{fig:mbu}(a). Since short-range interaction breaks the mean-field solvability of the Dicke model \cite{Zhu2019}, we also present results of TWA for finite $g_a$ in Fig.~\ref{fig:mbu}(b). Beyond mean-field results reveal the metastability of the TC phase for strong contact interaction as evident from the decaying envelope in the dynamics for $E_\mathrm{int}/E_\mathrm{rec}=0.10$. Nevertheless, the incommensurate TC remains stable against short-range interactions up to $E_\mathrm{int}/E_\mathrm{rec}=0.05$. Note that this $E_\mathrm{int}$ corresponds to a healing length of $\xi_a \equiv \hbar/\sqrt{2mg_a n_a} \approx 0.7\lambda_p $, where $m$ is the mass of an atom and $n_a$ is the initial density of the BEC. Our results suggest that the presence of weak short-range interaction will shift the phase boundaries in Fig.~\ref{fig:schem}(d) to higher pump strength, which means that the resonant frequency required to enter the TC phase is also increased.

\begin{figure}[!htb]
\centering
\includegraphics[width=1.0\columnwidth]{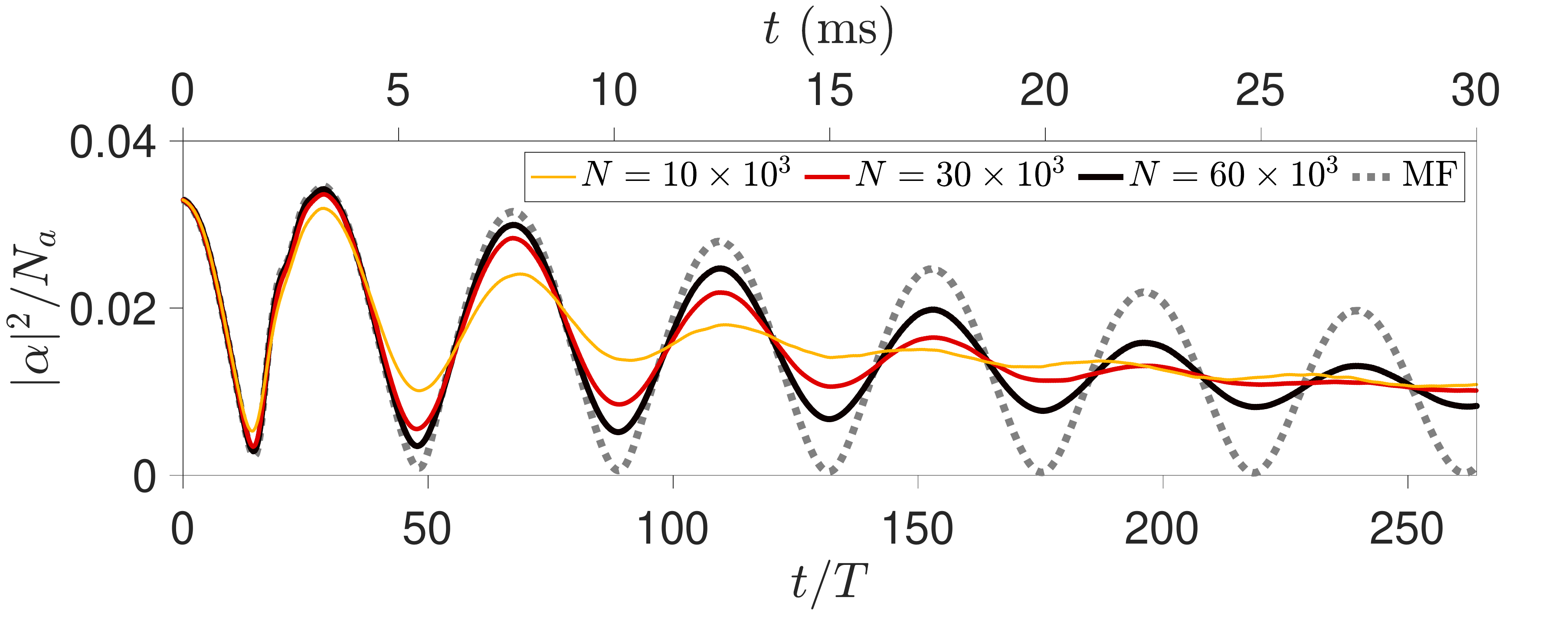}
\caption{Comparison of the cavity mode dynamics in the TC phase between the MF results for the thermodynamic limit $N_a\to \infty$ and the TWA results for finite $N_a$  (from yellow [light gray] to black for increasing $N_a$). Similar to Fig.~\ref{fig:mbu}, a Gaussian filter is also applied here to remove fast oscillations. The parameters are the same as Fig.~\ref{fig:power} except for $\Delta_0$, which is adjusted such that $N_a \Delta_0 = -2\pi \times 21.6~\mathrm{kHz}$ for varying $N_a$.}
\label{fig:finiteN} 
\end{figure} 
The dynamics for finite $N_a$ is amenable to TWA for $N_a/M \gg 1$, where $M$ is the number of single-particle modes relevant to the dynamics \cite{Polkovnikov2010,Blakie2008}. For the TC phase, we find that $M=81$ momentum modes are enough to guarantee numerical convergence and, thus, the condition $N_a/M \gg 1$ is safely satisfied for $N_a > 10^3$. The dynamics for finite but large $N_a$ are shown in Fig.~\ref{fig:finiteN}. The subharmonic oscillations are expected to decay for finite-sized dissipative systems but they should persist in the thermodynamic limit signalling the emergence of a genuine TC \cite{Iemini2018,Gambetta2018,Tucker2018}. In Fig.~\ref{fig:finiteN}, we show that indeed the oscillations decay for finite $N_a$ but their lifetime increases with $N_a$, which leads to an infinite decay time in the thermodynamic limit as predicted by our MF results.

\section{Conclusion}\label{sec:conc}

In this work, we have demonstrated the emergence of a time crystal in a shaken atom-cavity system. This dynamical phase breaks the space-translation symmetry of the optical dipole potential and the time-translation symmetry induced by the external shaking of the pump lattice. Spatial symmetry is broken by the formation of BDW$_1$ states, where atoms self-organize at the bonds between the antinodes of the light field. We find that exciting this new set of symmetry broken states, which does not exist in the equilibrium, also breaks time-translation symmetry as the system switches between BDW$_1$ states. 
The subharmonic frequency can be tuned by changing the driving frequency or the pump intensity, which suggests possible realization of both commensurate and incommensurate TC with a single dominant frequency.
We have constructed the dynamical phase diagram showing the transition into the TC phase and how it melts from a long-lived metastable state. 
We have shown that this phase possesses the general features of a TC, such as persistence of symmetry-breaking dynamics of an observable and rigidity to many-body correlations and perturbations in the driving protocol and noise sources inherent to the system. 
Our work opens up the possibility of using the atom-cavity platform to realize the simplest yet macroscopic incommensurate TC and further explore TTSB in driven-dissipative many-body systems.
Finally, we would like to emphasize that our findings are experimentally relevant as we have used physical parameters based on an existing atom-cavity setup in Ref.~\cite{Klinder2015}. Experimentally, the TTSB phenomenon predicted here can be observed \textit{in situ} from the dynamics of the cavity mode population, which oscillates at a much longer period than the external driving.

\acknowledgments

We acknowledge support from the Deutsche Forschungsgemeinschaft through the SFB 925 and the Hamburg Cluster of Excellence Advanced Imaging of Matter (AIM). We also thank Andreas Hemmerich and Louis-Paul Henry for useful discussions.  

\bibliography{biblio}

\end{document}